\documentclass[showpacs,amsmath,amssymb,pre,preprint]{revtex4}
\usepackage{graphicx}
 
\begin{document}

\title{An introduction to quantum plasmas}

\author{F. Haas}
\affiliation{Departamento de F{\'i}sica, Universidade Federal do Paran\'a, 81531-990, Curitiba, Paran\'a, Brazil}

\begin{abstract}
Shielding effects in non-degenerate and degenerate plasmas are compared. A detailed derivation of the Wigner-Poisson system is provided for electrostatic quantum plasmas where relativistic, spin and collisional effects are not essential. Later a detailed derivation of a quantum hydrodynamic model starting from the Wigner-Poisson system is shown. The route for this derivation considers the eikonal decomposition of the one-body wavefunctions of the quantum statistical mixture. The merits and limitations of the resulting quantum hydrodynamic model are discussed. 
\end{abstract}


\maketitle

\section{Introduction}
Plasma physics is frequently not a part of the standard physics curriculum. Thus, theoretical physicists who study quantum mechanics are not usually exposed to plasma science and have no interest in it. And vice versa, plasma physicists often do not have the required background in quantum theory. 

Plasma physics is commonly considered to be a purely classical field. 
However, in the last ten years there has been a renewed interest on plasma systems where quantum effects are important, see \cite{Haas}--\cite{Manfredi} for extensive reviews. Quantum me\-cha\-nics becomes relevant in plasmas when the de Broglie wavelength of the charge carriers is comparable to the inter-particle distance, so that there is a significant overlap of the corresponding wavefunctions. In such a si\-tua\-tion, the many-body problem is described by Fermi-Dirac statistics, in contrast to the usual laboratory and space plasmas which obey Maxwell-Boltzmann statistics. The fermionic character becomes prominent for sufficiently dense plasmas. Among these, one can cite plasmas in compact astrophysical objects such as in white dwarfs and the atmosphere of neutron stars \cite{Chabrier} or in the next ge\-ne\-ra\-tion intense laser-solid density plasma interaction experiments \cite{Marklund}. Moreover, many-body charged particle systems can not be treated by pure classical physics when the characteristic dimensions become comparable to the de Broglie wavelength. This is the case for quantum semiconductor devices, like high-electron-mobility transistors, resonant tunneling diodes or superlattices. The operation of these ultra-small devices rely on quantum tunneling of charge carriers through potential barriers. For these systems both Maxwell-Boltzmann or Fermi-Dirac statistics can be applied, according to the particle density. For ins\-tan\-ce, the Fermi-Dirac character should be taken into account in the drain region of $n^{+}nn^{+}$ diodes \cite{Marko, Jungel}. In addition, due to recent advances in femtosecond pump-probe spectroscopy, quantum plasma effects are attrac\-ting attention in the physics of metallic na\-no\-struc\-tu\-res and thin metal films \cite{Crouseilles}. Also, X-ray Thomson scattering in high energy density plasmas provide experimental techniques for accessing narrow bandwidth spectral lines \cite{Glenzer}, so as to detect frequency shifts due to quantum effects. Such modifications of the plasmon dispersion relation are inline with theoretical predictions \cite{Haas2}. In this context, precise experiments have been suggested \cite{Gregori} in order to measure low-frequency collective oscillations (ion-acoustic waves) in dense plasmas, as soon as keV free electron lasers will be available. 

The purpose of this short introduction is two-fold. First, we present an elementary discussion on the screening of a test charge in non-degenerate and de\-ge\-ne\-ra\-te plasmas. Shielding is one of the most emblematic collective effects in plasmas, so that it is worth to comment on the differences between the classical and quantum regimes of it. In this way a convenient introduction to quantum plasmas is provided. Second, we consider the basic kinetic and fluid models for quantum plasmas.  Hence we present the Wigner-Poisson system, which is the quantum equivalent for the Vlasov-Poisson system in classical plasma physics. Afterwards we discuss the derivation of fluid equations from the Wigner-Poisson model. Special attention is paid to the transition from microscopic to macroscopic des\-crip\-tions. While none of these subjects is new, most of the calculations here are intended to be shown in more detail than usually in the literature. In order to restrict the treatment to a basic level, only electrostatic, weakly coupled and non-relativistic quantum plasmas are con\-si\-de\-red. Hence, spin, magnetic field, collisional or relativistic effects are not addressed.  

This work is organized as follows. In Section II the basic physical parameters for non-degenerate and degenerate plasmas are introduced, by means of a discussion of the properties of shielding in these systems. Section III introduce the basic kinetic model for electrostatic quantum plasmas, namely the Wigner-Poisson system. Some of the basic properties of the Wigner function are analyzed, the Wigner function playing the r\^ole of a (quasi)-probability distribution in phase space. Section IV apply an eikonal, or Madelung decomposition of the one-body wavefunctions defining the quantum statistical ensemble. In this way the pressure functional is shown to be the sum of two terms associated to kinetic and osmotic velocity dispersion (to be defined later) plus a Bohm potential term associated to quantum diffraction effects. Section V is dedicated to final remarks. 

\section{Shielding in non-degenerate and degenerate plasmas}
Suppose a test charge $q_t > 0$ added to a plasma composed by an electron gas (of number density $n({\bf r})$) and a fixed homogeneous ionic background (of number density $n_0$). Initially, due to the Coulomb force the electrons would have trajectories deviated toward the test charge. Eventually, in the equilibrium situation a stationary cloud of negative charge would accumulate around $q_t$. Then, instead of $q_t$ an external observer would see an effective, smaller shielded charge. 
This is the (static) screening, or shielding effect in plasmas, which is in itself a manifestation of the quasi-neutrality property: due to the electric force, any excess charge tend to be compensated. In the stationary regime, the electrostatic field $\phi = \phi({\bf r})$ is described by Poisson's equation, 
\begin{equation}
\label{tc}
\nabla^{2}\phi = \frac{e}{\varepsilon_0}\,(n({\bf r}) - n_0) - \frac{q_t}{\varepsilon_0}\,\delta({\bf r}) \,,
\end{equation}
where $-e < 0$ is the electron charge, $\varepsilon_0$ is vacuum's permittivity and for definiteness the test charge is put at the origin. For simplicity we assume the test charge to be massive enough, so that it can be considered at rest. 

Assuming a Maxwell-Boltzmann statistics where the quiescent electron gas is in thermodynamic equilibrium at a temperature $T$, one would have 
\begin{equation}
\label{meq}
n({\bf r}) = n_0 \exp\left(\frac{e\phi}{\kappa_B T}\right) \,,
\end{equation}
where $\kappa_B$ is Boltzmann's constant. 

The Maxwell-Boltzmann statistics is appropriated for dilute, or non-degenerate plasmas, for which the degeneracy parameter $\chi = T_{F}/T \ll 1$. Here $T_F = E_F/\kappa_B$ is the Fermi temperature, defined \cite{Salinas} in terms of the Fermi energy $E_F$, given by 
\begin{equation}
\label{ef}
E_F = \frac{\hbar^2}{2m} (3\pi^2 n_0)^{2/3} \,, 
\end{equation}
where $\hbar$ is Planck's constant over $2\pi$ and $m$ is the electron mass. Since electrons are fermions (of spin $1/2$), even in the limit of zero thermodynamic temperature it is not possible to accommodate all of them in the ground state, due to the Pauli exclusion principle. Hence, excited states are filled up until the highest energy level, whose corresponding energy is defined as $E_F$.

Assuming the scalar potential to be zero before the insertion of the test charge, we can linearize Eq. (\ref{tc}) to obtain 
\begin{equation}
\label{ttc}
\nabla^{2}\phi = \frac{n_0 e^2}{\varepsilon_0 \kappa_B T}\,\phi - \frac{q_t}{\varepsilon_0}\,\delta({\bf r}) \,.
\end{equation}
The radially symmetric solution to Eq. (\ref{ttc}) with appropriate boundary conditions is the Yukawa potential
\begin{equation}
\label{scr}
\phi = \frac{q_t}{4\pi\varepsilon_0 r}\,e^{-r/\lambda_D} \,,
\end{equation}
where
\begin{equation}
\lambda_D = \left(\frac{\varepsilon_0\kappa_B T}{n_0 e^2}\right)^{1/2}
\end{equation}
is the (electron) Debye length. From Eq. (\ref{scr}), an observer would measure a very small potential for distances larger than the Debye length, which is inline with the linearization procedure.  Therefore $\lambda_D$ is a fundamental length for dilute, non-degenerate plasmas, playing the r\^ole of an effective range of the Coulomb interaction. The shielding effect is a collective effect due to a large number of electrons around $q_t$.  

What happens in the degenerate case? To start answering, notice that the number density (\ref{meq}) follows from the zeroth-order moment of the local Maxwell-Boltzmann one-particle equilibrium distribution function $f_{\rm cl} = f_{\rm cl}({\bf r},{\bf v})$ given by 
\begin{equation}
\label{eeq}
f_{\rm cl}({\bf r},{\bf v}) = n_0 \left(\frac{m}{2\pi\kappa_B T}\right)^{3/2} \exp\left[- \frac{1}{\kappa_B T}\left(\frac{m v^2}{2} - e\phi\right) \right] \,. 
\end{equation}
In other words, 
\begin{equation}
\label{cuf}
n({\bf r}) = \int d{\bf v} \,f_{\rm cl}({\bf r},{\bf v})  \,.
\end{equation}

On the other hand, the simplest approach for a degenerate plasma would consider an uniform distribution of electrons for energy smaller than the Fermi energy, and no particles above the Fermi level. Then Eq. (\ref{eeq}) could be replaced by 
\begin{equation}
\label{tfu}
f_{\rm cl}({\bf r},{\bf v}) = \frac{3 n_0}{4\pi v_F^3} \quad {\rm if} \quad \frac{m v^2}{2} - e\phi < E_F 
\end{equation}
with $f_{\rm cl}({\bf r},{\bf v}) = 0$ otherwise. In Eq. (\ref{tfu}) we have the Fermi velocity $v_F = (2E_F/m)^{1/2}$. Notice the energy shift due to a non-zero scalar potential, in the same manner as for the local Maxwell-Boltzmann equilibrium given by Eq. (\ref{eeq}). The distribution (\ref{tfu}) is representative of a zero-temperature Thomas-Fermi \cite{fren} equilibrium, displaying equal occupation probabilities for energies smaller than Fermi's energy, and no particles beyond. At this point temperature effects are disregarded. 

Equation (\ref{cuf}) can be used to find the number density for the Thomas-Fermi equilibrium (\ref{tfu}), yielding 
\begin{equation}
\label{nfe}
n({\bf r}) = n_0 \left(1 + \frac{e\phi}{E_F}\right)^{3/2} \,.
\end{equation}
Inserting this result into Poisson's equation (\ref{tc}) and linearizing we get 
\begin{equation}
\label{tttc}
\nabla^{2}\phi = \frac{3 n_0 e^2}{2 \varepsilon_0 E_F}\,\phi - \frac{q_t}{\varepsilon_0}\,\delta({\bf r}) \,.
\end{equation}
which is the same as Eq. (\ref{ttc}) with the replacement $\kappa_B T \rightarrow E_F$, except for a numerical factor. Therefore, a shielding distance $\lambda_F$, or Thomas-Fermi length \cite{Shukla2}, can be set for degenerate plasmas,
\begin{equation}
\label{lf}
\lambda_F = \left(\frac{2\varepsilon_0 E_F}{3n_0 e^2}\right)^{1/2} \,.
\end{equation}
The Thomas-Fermi length is non-zero even for zero thermodynamic temperature, unlike the Debye length $\lambda_D$. This happens because of the exclusion principle which prevents the accumulation of electrons in the same place as the test charge.   

Notice that in a sense every particle in a plasma, be it degenerate or not, can be interpreted as a test charge with the corres\-pon\-ding screening cloud. Hence instead of the long-range Coulomb field we have an effective Yukawa interaction field. This point of view is adopted, for instance, in the treatment of the ultrafast phase-space dynamics of ultracold, neutral plasmas \cite{Murillo}. 

In our simplified picture $f_{\rm cl}({\bf r},{\bf v})$ was regarded as a purely classical probability distribution function. A more detailed treatment taking into account quantum diffraction effects shows that at large distance the inter-particle oscillation behave as
\begin{equation}
\phi \sim \frac{1}{r^3} \cos(2r/\lambda_F) \,,
\end{equation}
a phenomenon known as Friedel oscillations \cite{Friedel, Lind}. As discussed in the next Section, the Wigner function provides a convenient tool to incorporate quantum effects in plasmas, in strict analogy with the classical probability distribution approach. 

The analysis of shielding allows the introduction of  characteristic length scales $\lambda_D$ and $\lambda_F$ in non-degenerate and degenerate plasmas, respectively. In addition a pertinent length scale measuring the spatial extension of the wavefunction of each electron is the de Broglie wavelength $\lambda_B = \hbar/(m v_T)$,  where $v_T = (2\kappa_B T/m)^{1/2}$ is the thermal velocity. Given the de Broglie wavelength one can express the degeneracy parameter as 
\begin{equation}
\chi = \frac{T_F}{T} = \frac{1}{2} (3\pi^2 n_0 \lambda_{B}^{3})^{2/3} \,.
\end{equation}
Hence Fermi-Dirac statistics is necessary when $\lambda_B$ is of the same order of the inter-particle distance. 

Other fundamental scales in classical and quantum plasmas are as follows.
\begin{itemize}
\item Time scale for both classical and quantum plasmas: $\omega_{p}^{-1}$, where 
\begin{equation}
\omega_p = \left(\frac{n_0 e^2}{m \varepsilon_0}\right)^{1/2} \,.
\end{equation}
is the plasma frequency. Given some electron charge depletion, an electric force appears in order to restore complete charge neutrality. The resulting linear oscillations have a frequency $\omega_p$. 
\item Typical interaction energy $U_{\rm int}$ for both classical and quantum plasmas: 
\begin{equation}
U_{\rm int} = \frac{e^2 n_{0}^{1/3}}{\varepsilon_0} \,,
\end{equation}
since the mean inter-particle distance scales as $n_{0}^{-1/3}$. 
\item Typical kinetic energies: $K_C = \kappa_B T$ for non-degenerate and $K_Q = \kappa_B T_F$ for de\-ge\-ne\-ra\-te plasmas. This happens for fermions because of the filling up of excited states due to the Pauli exclusion principle, even at zero temperature. 
\end{itemize}

From the typical energy scales we can form classical $\Gamma_C$ and quantum $\Gamma_Q$ energy coupling parameters,  
\begin{eqnarray}
\label{gcc}
\Gamma_C &=& \frac{U_{\rm int}}{K_C} =  \frac{e^2 n_{0}^{1/3}}{\varepsilon_0 \kappa_B T} = 2.1 \times 10^{-4} \times \frac{n_{0}^{1/3}}{T} \,,\\
\label{gqq}
\Gamma_Q &=& \frac{U_{\rm int}}{K_Q} = \frac{2 m e^2}{(3\pi^2)^{2/3} \varepsilon_0 \hbar^2 n_{0}^{1/3}} = 5.0 \times 10^{10}\, n_{0}^{-1/3} \,.
\end{eqnarray}
Numerical values are for S. I. units. Weakly coupled plasmas have small energy coupling parameters. From Eqs. (\ref{gcc}--\ref{gqq}) the conclusion is that while classical weakly coupled plasmas tend to be dilute and cold, quantum weakly coupled plasmas tend to be dense. For example, with $n_0 > 10^{35} m^{-3}$ (white dwarf) one has $\Gamma_Q < 0.1$, allowing the use of collisionless models in a first approximation. This is a consequence of the Pauli exclusion principle, which contribute to forbids e-e collisions in very dense systems. In other words, the more dense a degenerate plasma is, the more it resembles an ideal gas, except for the mean field, collective interaction. 

A more detailed account on the cha\-rac\-te\-ris\-tic scales in classical and quantum Coulomb systems can be found in \cite{Haas, Manfredi, Bonitz}. 

\section{Obtaining the Wigner-Poisson system}
The Vlasov-Poisson system 
\begin{eqnarray}
\label{vx}
\frac{\partial f_{cl}}{\partial t} &+& v\frac{\partial f_{cl}}{\partial x} + \frac{e}{m}\frac{\partial\phi}{\partial x}\frac{\partial f_{cl}}{\partial v} = 0 \,,\\
\label{cuff}
\frac{\partial^2\phi}{\partial x^2} &=& \frac{e}{\varepsilon_0}\left(\int dv f_{cl}(x,v,t) - n_0\right) 
\end{eqnarray}
is the basic tool for kinetic theory of classical plasmas, 
where for simplicity we take an one-dimensional electron plasma in a fixed neutralizing ionic background $n_0$. In the Vlasov equation (\ref{vx}), $f_{cl} = f_{cl}(x,v,t)$ is the reduced one-particle probability distribution function. For a plasma with $N$ electrons,  
$(1/N) f_{cl}(x,v,t)\,dv\,dx$ gives the pro\-ba\-bi\-li\-ty of finding one electron with position between $x$ and $x+dx$ and velocity between $v$ and $v+dv$, at the time $t$. Hence, the normalization
\begin{equation}
\int\,dv\,dx\,f_{cl}(x,v,t) = N 
\end{equation}
is assumed. 

From the knowledge of $f_{cl}$, obtained solving either analytically or numerically the Vlasov-Poisson system subject to appropriate boundary conditions, the whole ma\-chi\-ne\-ry of classical statistical mechanics can be used to compute the expectation values of macroscopic quantities. For example, the a\-ve\-ra\-ge kinetic energy of one electron follows from
\begin{equation}
< \frac{m v^2}{2} > = \frac{1}{N} \int dv dx f_{cl}(x,v,t) \frac{m v^2}{2} \,.
\end{equation}
It is convenient to adopt a similar me\-tho\-do\-lo\-gy in a quantum kinetic theory for plasmas, as far as possible. 

Quantum mechanics can be formulated in phase-space using the Wigner function $f = f(x,v,t)$, which provides a quantum equivalent of $f_{cl}(x,v,t)$. Let us study  some of the basic properties of the Wigner function. For an one-particle pure state quantum system with wavefunction $\psi(x,t)$, the Wigner function is  defined \cite{Wign} by 
\begin{equation}
f = \frac{m}{2\,\pi\,\hbar}\int\,ds\,\exp\left(\,\frac{i\,m\,v\,s}{\hbar}\right)\,\psi^{*}\left(x+\frac{s}{2},t\right)\psi\left(x-\frac{s}{2},t\right) \,,
\end{equation}
with all symbols as before. From $f(x,v,t)$ we can compute the probability density 
\begin{equation}
\label{density}
\int\,dv\,f(x,v,t) = |\psi(x,t)|^2
\end{equation}
and the probability current
\begin{equation}
\label{cur}
\int\,dv\,f(x,v,t)\,v = \frac{i\,\hbar}{2\,m}\,\left(\psi\frac{\partial\psi^*}{\partial\,x} - \psi^{*}\frac{\partial\psi}{\partial\,x}\right) \,.
\end{equation}
The above wavefunction is normalized to unity. 

In general, even an one-particle quantum system can not be represented by a wavefunction alone. Mixed quantum states are described by a quantum statistical ensemble $\{\psi_{\alpha}(x,t)\,,p_{\alpha}\}, \alpha = 1, 2, ... M$, with each wavefunction $\psi_{\alpha}(x,t)$ having an occupation pro\-ba\-bi\-li\-ty $p_{\alpha}$ such that $p_{\alpha} \geq 0\,, \sum_{\alpha = 1}^{M}p_{\alpha} = 1$. The Wigner function is then defined by the superposition
\begin{equation}
\label{mixed}
f = \frac{m}{2\,\pi\,\hbar}\sum_{\alpha = 1}^{M}\,p_{\alpha}\,\int\,ds\,\exp\left(\,\frac{i\,m\,v\,s}{\hbar}\right)\,\psi^{*}_{\alpha}\left(x+\frac{s}{2},t\right)\psi_{\alpha}\left(x-\frac{s}{2},t\right) \,.
\end{equation}
Correspondingly, we have the probability and current densities 
\begin{eqnarray}
\label{n1}
\int\,dv\,f(x,v,t) &=& \sum_{\alpha = 1}^{M}\,p_{\alpha}\,|\psi_{\alpha}(x,t)|^2  \,,\\
\label{j1}
\int\,dv\,f(x,v,t)\,v &=& \frac{i\,\hbar}{2\,m}\,\sum_{\alpha = 1}^{M}\,p_{\alpha}\,\left(\psi_{\alpha}\frac{\partial\psi^{*}_{\alpha}}{\partial\,x} - \psi^{*}_{\alpha}\frac{\partial\psi_{\alpha}}{\partial\,x}\right)  \,. 
\end{eqnarray}

The density matrix $\rho(x,y,t)$ could also be elected as the central object in a quantum kinetic theory for plasmas. However, in this manner the similarity to the Vlasov-Poisson model would be lost. Besides, we have a complete correspondence since
\begin{eqnarray}
\rho(x,y,t) &\equiv& \sum_{\alpha = 1}^{M}\,p_{\alpha}\,\psi_{\alpha}(x,t)\,\psi_{\alpha}^{*}(y,t) \nonumber \\ 
\label{dm}
&=& \int\,dv\,\exp\left(\frac{i\,m\,v\,(x-y)}{\hbar}\right)\,f\left(\frac{x+y}{2},v,t\right) \,,
\end{eqnarray}
with inverse given by  
\begin{equation}
\label{wf}
f(x,v,t) = \frac{m}{2\,\pi\,\hbar}\int\,ds\,\exp\left(\frac{i\,m\,v\,s}{\hbar}\right)\,\rho\left(x+\frac{s}{2},x-\frac{s}{2},t\right) \,.
\end{equation}
In other words, given the Wigner function the density matrix can be found and vice-versa. The same apply to many-body Wigner functions and density matrices. 

Going one step further, consider now a $N-$particle statistical mixture described by the set $\{\psi_{\alpha}^{N}(x_{1},x_{2},...,x_{N},t)\,, p_{\alpha}\}$, where the normalized $N-$particle ensemble wavefunctions $\psi_{\alpha}^{N}(x_{1},x_{2},...,x_{N},t)$ are distributed with probabilities $p_{\alpha}\,, \alpha = 1,2,...,M$ sa\-tis\-fy\-ing $p_{\alpha} \geq 0, \sum_{\alpha = 1}^{M} p_{\alpha} = 1$ as before. Here $x_i$ re\-pre\-sents the position of the $i$th-particle, $i = 1,2,...N$. All particles have the same mass $m$. The $N-$particle Wigner function is defined by
\begin{eqnarray}
f^{N}(x_{1},v_{1}\!\!\!\!&,&\!\!\!\!...,x_{N},v_{N},t) = N\,\left(\frac{m}{2\,\pi\,\hbar}\right)^{N}\sum_{\alpha = 1}^{M}\,p_{\alpha}\,\int\,ds_{1}...ds_{N}\,\exp\left(\frac{i\,m\,\sum_{i=1}^{N}\,v_{i}s_{i}}{\hbar}\right)  \nonumber \\ 
\label{nbody}
&\times& \,\psi^{N\,*}_{\alpha}\left(x_{1}+\frac{s_1}{2},...,x_{N}+\frac{s_N}{2},t\right)\,\psi_{\alpha}^{N}\left(x_{1}-\frac{s_1}{2},...,x_{N}-\frac{s_N}{2},t\right) \,.
\end{eqnarray}
The factor $N$ in Eq. (\ref{nbody}) is inserted so that 
\begin{equation}
\int dx_{1}dv_{1}...dx_{N}dv_{N}f^{N}(x_{1},v_{1},...,x_{N},v_{N},t) = N .
\end{equation}
In this manner, the integral of $f^N$ over all the velocities gives a number density in con\-fi\-gu\-ra\-tion space. The fermionic character of the system is not necessarily included, but this could be done assuming the $N-$body ensemble wavefunctions to be antisymmetric in Eq. (\ref{nbody}).   

Wigner functions provide a convenient mathematical tool to calculate average quantities, in the same way as for the classical probability distribution function. However, they are not necessarily positive definite. Hence it is customary to refer to them as quasi-probability distributions. 

To proceed to the derivation of the quantum analog of the Vlasov-Poisson system (\ref{vx}--\ref{cuff}), we can introduce the reduced one-particle Wigner function $f(x_{1},v_{1},t)$,  
\begin{equation}
\label{oneb}
f(x_{1},v_{1},t) = \int\,dx_{2}dv_{2}...dx_{N}dv_{N} f^{N}(x_{1},v_{1},...,x_{N},v_{N},t) \,,
\end{equation}
and the reduced two-particle Wigner function $f^{(2)}(x_{1},v_{1},x_2,v_2,t)$ with a convenient nor\-ma\-li\-za\-tion factor $N$,
\begin{equation}
\label{f2}
f^{(2)}(x_{1},v_{1},x_2,v_2,t) = N\,\int\,dx_{3}dv_{3}...dx_{N}dv_{N}\,f^{N}(x_{1},v_{1},...,x_{N},v_{N},t) \,.
\end{equation}
Similar reduced $N-$particle Wigner functions with $N \geq 3$ can be likewise defined. 
If the Wigner function were a true probability distribution, $(1/N)\,f(x_1,v_1,t) dx_1 dv_1$ would give the probability of finding the particle $1$ in an area $dx_1 dv_1$  centered at $(x_1,v_1)$, irres\-pec\-ti\-ve of the ``position" and "velocity" of the remaining $i$th-particles, $i = 2,...,N$. An analogous ``classical" probability interpretation could be assigned to the remaining reduced Wigner functions, except for the non-positive-definiteness property. 

In passing, we have 
\begin{equation}
\int dx_1 dv_1 f(x_{1},v_{1},t) = N \,, \quad \int dx_1 dv_1 dx_2 dv_2 f^{(2)}(x_{1},v_{1},x_2,v_2,t) = N^2 \,.
\end{equation}

It is reasonable to pay special attention to the reduced Wigner functions, since $f^N$ contain far more information than what is commonly needed. In this regard the one-particle Wigner function plays a distinguished r\^ole. Indeed, macroscopic objects like number and current densities can be derived from $f$ after integration over just one velocity variable, exactly as in Eqs. (\ref{n1}--\ref{j1}), originally written for an one-particle system. 

To obtain the evolution equation satisfied by the one-body Wigner function, the phi\-lo\-so\-phy of Ref. \cite{Klim} can be pursued. Consider then the Schr\"odinger equation satisfied by the $N-$body ensemble wavefunctions, 
\begin{equation}
\label{sn}
i\hbar\frac{\partial\psi_{\alpha}^N}{\partial t} = - \frac{\hbar^2}{2m}\,\sum_{i=1}^{N}\frac{\partial^2\psi_{\alpha}^N}{\partial\,x_{i}^2} + V(x_{1},...,x_{N})\,\psi_{\alpha}^{N}
\end{equation}
for an interaction energy $V(x_{1},...,x_{N})$. 

We are concerned with the case where the system components interact through some two-body potential $W$, 
\begin{equation}
\label{cou}
V(x_{1},...,x_{N}) = \sum_{i<j}\,W(|x_i - x_j|) \,.
\end{equation}
This situation is evidently interesting because of the Coulomb force. 

Some algebra shows \cite{Haas, Klim} that  
\begin{eqnarray}
\label{1b}
\frac{\partial\,f}{\partial\,t} &+&
v_1\,\frac{\partial\,f}{\partial\,x_1} = - \frac{i\,m}{2\pi\hbar^2}\,\int\,ds_1\,dv_1'\,dx_2\,dv_2'\,\exp\left(-\frac{im(v_1'-v_1)\,s_1}{\hbar}\right) \\ 
&\times& \left(W(|x_1-x_2+\frac{s_1}{2}|) - W(|x_1-x_2-\frac{s_1}{2}|)\right)\,f^{(2)}(x_1,v_{1}',x_2,v_2',t) \,. \nonumber 
\end{eqnarray}
in terms of the reduced two-particle Wigner function $f^{(2)}$. In the derivation, $N \gg 1$ was taken into account. Actually a more detailed argument in\-vol\-ving the higher-order Wigner functions yield a quantum BBGKY (Bogoliubov-Born-Green-Kirkwood-Yvon) hie\-rar\-chy \cite{Bogoliubovb, Born, Kirkwoodb, Yvon}, where the dynamics of the $(N-1)$-body reduced Wigner function is shown to depend on the $N$-body reduced Wigner function. Hence in both the classical infinite BBGKY set of equations and its quantum analogue we are faced with a closure problem. 

Ignoring correlations is the simplest way to close the system, considering that the distribution of particles at $(x_i,v_i)$ is not affected by particles at a distinct phase space point $(x_j,v_j)$. In this mean field (or Hartree) approximation the $N-$body Wigner function factorizes,
\begin{equation}
f^{(2)}(x_1,v_1,x_2,v_2,t) = f(x_1,v_1,t)\,f(x_2,v_2,t) \,.
\end{equation}
Equation (\ref{1b}) then becomes 
\begin{equation}
\frac{\partial\,f}{\partial\,t} +
v_1\,\frac{\partial\,f}{\partial\,x_1} = \int\,dv_1'\,K[W_{sc}\,|\,v_1' - v_1,x_1,t]\,f(x_1,v_1',t) \,,  
\end{equation}
with the mean field self-consistent potential 
\begin{equation}
\label{av}
W_{sc}(x,t) = \int\,dv\,dx'\,f(x',v,t)\,W(|x - x'|) \,.
\end{equation}
The functional $K[W_{sc}\,|\,v_1' - v_1,x_1,t]$ is defined by
\begin{eqnarray}
K[W_{sc}\,|\,v_1' - v_1,x_1,t] &=& - \frac{im}{2\pi\hbar^2}\int\,ds_1\,
\exp\left(- \frac{im(v_1' - v_1)s_1}{\hbar}\right) \times \nonumber  \\
&\times& \left(W_{sc}(x_1 +
\frac{s_1}{2},t) - W_{sc}(x_1 - \frac{s_1}{2},t)\right) \,.
\end{eqnarray}

Frequently an external, possibly time-dependent potential $V_{ext}(x_1,...,x_N,t)$ should be included. For instance, such a circumstance arises in solid state devices, when considering the electronic motion in a fixed ionic lattice or under a confining field like in quantum wires or quantum wells \cite{Marko, Jungel, DiVentra}. Or even the field due to an homogeneous ionic background can be thought as an external superimposed field. Hence, consider an external potential of the form 
\begin{equation}
\label{exte}
V_{ext}(x_1,...,x_N,t) = \sum_{i=1}^{N}\,W_{ext}(x_i,t) 
\end{equation}
for some one-particle potential $W_{ext}(x_i,t)$. Implicitly in Eq. (\ref{exte}), the functional form of $W_{ext}$ is the same irrespective of $x_i$, implying that the external field has the same influence on all particles. For completeness we indicate the changes for a potential 
\begin{equation}
V(x_{1},...,x_{N}) = \sum_{i<j}\,W(|x_i - x_j|) + \sum_{i=1}^{N}\,W_{ext}(x_i,t) \,.
\end{equation}
Following the same steps as before, the one-body reduced Wigner function $f(x_1,v_1,t)$ can then be shown to satisfy
\begin{equation}
\label{weq}
\frac{\partial\,f}{\partial\,t} +
v_1\,\frac{\partial\,f}{\partial\,x_1} = \int\,dv_1'\,K[W_{sc} + W_{ext}\,|\,v_1' - v_1,x_1,t]\,f(x_1,v_1',t) \,,  
\end{equation}
for
\begin{eqnarray}
K[W_{sc} &+& W_{ext}\,|\,v_1' - v_1,x_1,t] = - \frac{im}{2\pi\hbar^2}\int\,ds_1\,
\exp\left(- \frac{im(v_1' - v_1)s_1}{\hbar}\right) \times \nonumber  \\
&\times& \left(W_{sc}(x_1 +
\frac{s_1}{2},t) + W_{ext}(x_1 + \frac{s_1}{2},t)  - W_{sc}(x_1 - \frac{s_1}{2},t) - W_{ext}(x_1 - \frac{s_1}{2},t)\right) \nonumber \\
\end{eqnarray}
and for the averaged self-consistent potential $W_{sc}$ the same as in Eq. (\ref{av}). 

The necessary changes in three-dimensional charged particle motion are as follows. Assume the Coulomb interaction 
\begin{equation}
W(|{\bf r} - {\bf r}'|) = \frac{e^2}{4\pi\varepsilon_0\,|{\bf r} - {\bf r}'|} \,. 
\end{equation}
and define the total electrostatic potential $\phi({\bf r},t)$ so that
\begin{equation}
\phi({\bf r},t) = \phi_{sc}({\bf r},t) + \phi_{ext}({\bf r},t) \,,
\end{equation}
in terms of the self-consistent $W_{sc}$ and some external $W_{ext}$ potentials, 
where 
\begin{equation}
W_{sc}({\bf r},t) = - e\phi_{sc}({\bf r},t) \,, \quad W_{ext}({\bf r},t) = - e\phi_{ext}({\bf r},t) \,.
\end{equation}

From the three-dimensional version of Eq. (\ref{av}) it follows that 
\begin{eqnarray}
\nabla^2\phi_{sc} &=& - \frac{e}{\varepsilon_0}\int\,d{\bf v}\,d{\bf r}'\,f({\bf r}',{\bf v},t)\,\nabla^{2}\left(\frac{1}{4\pi\,|{\bf r} - {\bf r}'|}\right) \nonumber \\ 
&=& \frac{e}{\varepsilon_0}\int\,d{\bf v}\,d{\bf r}'\,f({\bf r}',{\bf v},t)\,\delta({\bf r} - {\bf r}') \nonumber \\ \label{sco}
&=& \frac{e}{\varepsilon_0}\int\,d{\bf v}\,f({\bf r},{\bf v},t) \,.
\end{eqnarray}
Moreover, 
\begin{equation}
\label{exp}
\nabla^2\phi_{ext} = - \frac{1}{e}\nabla^{2}W_{ext} \equiv - \frac{n_0 e}{\varepsilon_0} 
\end{equation}
if the external potential is due to an immobile fixed homogeneous ionic background of density $n_0$ and ion charge $e$. Appropriate changes \cite{Marko, Jungel, DiVentra} are needed in the case of a non-homogeneous background \textit{e.g.} as in the case of doped semiconductors, or in the presence of a dispersive medium with a permittivity constant $\varepsilon \neq \varepsilon_0$. 

From Eqs. (\ref{sco}--\ref{exp}) it is immediate to derive
\begin{equation}
\nabla^2\phi = \frac{e}{\varepsilon_0}\left(\int\,d{\bf v}\,f({\bf r},{\bf v},t) - n_0\right) \,,
\end{equation}
the Poisson equation in this case. 

For notational simplicity, it is indicated to restrict again to the one-dimensional case. In terms of the electrostatic potential $\phi$, Eq. (\ref{weq}) is rephrased as 
\begin{equation}
\label{w1}
\frac{\partial\,f}{\partial\,t} +
v\,\frac{\partial\,f}{\partial\,x} = \int\,dv'\,K_{\phi}[\phi\,|\,v' - v,x,t]\,f(x,v',t) \,, 
\end{equation}
where $K_{\phi}[\phi\,|\,v' - v,x,t]$ is the functional
\begin{eqnarray}
K_{\phi}[\phi\,|\,v' - v,x,t] &=& \frac{iem}{\hbar}\int\frac{ds}{2\pi\hbar}
\exp\left(\frac{im(v' - v)s}{\hbar}\right) \times \nonumber  \\
\label{w3}
&\times& \left(\phi(x +
\frac{s}{2},t) - \phi(x - \frac{s}{2},t)\right) \,.
\end{eqnarray}
Equation (\ref{w1}) can be termed the quantum Vlasov equation (in the electrostatic case), since it is the quantum analog of the Vlasov equation satisfied by the reduced one-particle probability distribution function. Finally, the quantum Vlasov equation should be coupled to Poisson's equation, 
\begin{equation}
\label{w2}
\frac{\partial^{2}\phi}{\partial\,x^2}
= \frac{e}{\varepsilon_0}\left(\int\,dv\,f(x,v,t) - n_{0}\right) \,. 
\end{equation}

Equations (\ref{w1}) and (\ref{w2}) constitute the Wigner-Poisson system, which is the fundamental kinetic model for electrostatic quantum plasmas. It determines in a self-consistent way both the Wigner function, associated to how the particles distribute in phase space, and the scalar potential, which in turn describe the forces acting on the particles. 

The Wigner-Poisson system needs to be supplemented with suitable boundary and initial conditions. For plasmas, decaying or periodic boundary conditions are frequently employed. For nano-devices, the choice of boundary conditions is subtler due to the finite size of the system and the nonlocal character of the Wigner function. Indeed, to compute the integral defining the Wigner function we need to specify $f(x,v,0)$ in the whole space even when dealing with finite size systems \cite{Marko, Jungel, DiVentra}. 

Let us review the necessary steps for the derivation of the Wigner-Poisson system. Above all, it is a mean field model with the $N-$body ensemble Wigner function supposed to be factorisable, in order to achieve the simplest closure of the quantum BBGKY hierarchy. Hence the $N-$body Schr\"odinger equation (or the Liouville-von Neumann equation for the $N-$body ensemble density matrix) is replaced by a system with fewer degrees of freedom. Indeed, for $N$ electrons in three-dimensional space, we have $3N+1$ coordinates to define the wavefunction, $6N+1$ coordinates for the density matrix, and only $6+1 = 7$ independent variables for the reduced one-body Wigner function, here taking into account time. Therefore, the mean field theory is much less numerically demanding, since it deserve the discretization of a space with fewer dimensions. The price for the reduction is the neglect of collisions, besides spin and relativistic effects at least in the present formulation. Finally, no magnetic fields were included. 

Since it is the analog to the Vlasov-Poisson, the Wigner-Poisson system becomes the natural tool in quantum kinetic theory for electrostatic plasmas. This is because the methods applied to the Vlasov-Poisson system can with some optimism be translated to Wigner-Poisson quantum plasmas. Nevertheless, other quantum kinetic treatments for charged particle systems are obviously important. For example, the density functional \cite{DiVentra} and Green's function \cite{Haug, Kadanoff} approaches are popular tools for the modeling of quantum transport among the condensed matter community. Moreover, the simplifications of the Wigner-Poisson model can be sometimes overcomed by alternative formulations. For instance, Green's function techniques can be used to describe collisions associated to short range particle-particle interactions \cite{Haug, Kadanoff}, in terms of a  Boltzmann type collision operator.  

It is interesting to look to the semiclassical limit of the quantum Vlasov equation (\ref{w1}). By means of the change of variable $s = \hbar\,\tau/m$ and Taylor expanding, one get
\begin{equation}
\label{w4}
\frac{\partial\,f}{\partial\,t} +
v\,\frac{\partial\,f}{\partial\,x} + \frac{e}{m}\,\frac{\partial\phi}{\partial x}\frac{\partial\,f}{\partial\,v} =  \frac{e\hbar^2}{24\,m^3}\frac{\partial^3\phi}{\partial\,x^3}\,\frac{\partial^3\,f}{\partial v^3} + O(H^4) \,.
\end{equation}
Implicitly, the semiclassical approximation assumes the smallness of a non-dimensional quantum parameter $H = \hbar/(m v_0 L_0)$, where $v_0$ and $L_0$ are resp. characteristic velocity and length scales. 

Equation (\ref{w4}) is a semiclassical Vlasov equation, with $f$ playing the r\^ole of one-particle distribution function. We see that unlike for classical plasmas, in general neither $f$ nor phase space volume are preserved by the quantum Vlasov equation, since 
\begin{equation}
\label{bol}
\frac{df}{dt} =  \frac{e\hbar^2}{24m^3}\frac{\partial^3\phi}{\partial\,x^3}\,\frac{\partial^3\,f}{\partial v^3} + O(H^4) \neq 0
\end{equation}
along the (classical) characteristic equations 
\begin{equation}
\frac{dx}{dt} = v \,, \quad \frac{dv}{dt} = \frac{e}{m}\frac{\partial\phi}{\partial x} \,.
\end{equation}
Due to this property the positive de\-fi\-ni\-te\-ness of the Wigner function is not preserved by Eq. (\ref{w1}), except for linear electric fields and a vanishing quantum correction. Also notice that Eq. (\ref{bol}) is a Boltzmann's like equation, although this is not exactly true since one has time-reversal invariance under $t \rightarrow - t, x \rightarrow x, v \rightarrow - v, f(x,v,t) \rightarrow f(x,-v,-t)$ (assuming $\phi(x,t) = \phi(x,-t)$). There is no irreversibility nor memory loss at all in the quantum Vlasov equation, be it in the semiclassical or the fully quantum versions. This is not exactly surprising since the Schr\"odinger equation is time-reversal invariant. 

Even when the quantum Vlasov and Vlasov equations are the same, which happens for linear electric fields,  $f(x,v,t)$ can not be considered as an ordinary probability distribution function. Not all functions on phase space can be taken as Wigner functions, since a genuine Wigner function necessarily correspond to a positive definite density matrix. Hence, we have at least \cite{Hillery} the following necessary conditions, 
\begin{eqnarray}
\label{cw1}
\int dx dv f &=& N \,,\\
\label{cw3}
\int dv f &\geq& 0 \,,\\
\label{cw4}
\int dx f &\geq& 0 \,,\\
\label{cw2}
\int dx dv f^2 &\leq& \frac{m\,N^2}{2\pi\hbar} \,.
\end{eqnarray}
Equation (\ref{cw1}) is simply a normalization condition, while Eqs. (\ref{cw3}) and (\ref{cw4}) assure the spatial and velocity marginal probability densities to be everywhere non-negative. Equation (\ref{cw2}) eliminate too spiky Wigner functions, which would be against the uncertainty principle.  
For instance, for the Gaussian profile
\begin{equation}
f = \frac{N}{2\pi\sigma_x\sigma_v}\,\exp\left(-\frac{x^2}{2\sigma_{x}^2}\right)\,\exp\left(-\frac{v^2}{2\sigma_{v}^2}\right) 
\end{equation}
with constant standard deviations $\sigma_{x,v}$ from Eq. (\ref{cw2}) it follows that
\begin{equation}
\sigma_x \sigma_v > \frac{\hbar}{2m} \,,
\end{equation}
in accordance with the uncertainty principle. 

Since no collisional effects are included in the Wigner-Poisson system, strongly coupled quantum plasmas deserves special treatment. Hence in principle the Wigner-Poisson model assumes a small energy coupling parameter, see Eqs. (\ref{gcc}--\ref{gqq}). However, for dense plasmas (as the electron gas in metals) sometimes quantum collisionless models are still applicable thanks to the Pauli blocking phenomenon preventing e-e collisions \cite{Manfredi}, even if the energy coupling parameter is large. 
 
Not only very dense charged particle systems deserve quantum kinetic equations. For instance, due to the ongoing miniaturization, even scarcely populated electronic systems such as resonant tunneling diodes \cite{Marko} should be described in terms of quantum models. Indeed, the behavior of these ultra-small electronic devices relies on quantum diffraction effects as tunneling, making purely classical methods inappropriate. The non-local integro-differential potential term in Eq. (\ref{w1}) in the Wigner-Poisson system has been shown to be capable of the modeling of negative differential resistance, associated to tunneling \cite{Kluk}. Moreover, the collisionless approximation becomes more reasonable in view of the nanometric scale of the devices, simply because the mean free-path exceeds the system size. In the same manner, the usually extreme high operating frequencies makes the collisionless approximation more accurate, because $\omega\tau \ll 1$ for an operating frequency $\omega$ and a average time $\tau$ between collisions. For example, in resonant tunneling diodes one can find \cite{Marko} potential barriers of the order $0.3\, eV \sim \hbar\omega$, implying an operating frequency $\omega \sim 10^{15}\, s^{-1}$. Therefore the Wigner-Poisson system is well suited for ballistic, collisionless processes in nanometric solid state devices, even at relatively low densities $n_0 \sim 10^{24} m^{-3}$. Correspondingly one finds a Fermi temperature $T_F \sim 40 K$ much smaller than a typical room temperature $T \sim 300 K$, justifying the non-degeneracy assumption and Maxwell-Boltzmann's statistics.   

\section{Derivation of a fluid model for quantum plasmas}
The Wigner-Poisson method presents
some drawbacks: (a) it is a nonlocal, integro-differential system;
(b) its numerical treatment requires the discretization of 
the whole phase space.
Moreover, as is often the case with kinetic models, the Wigner-Poisson system
gives more information than one is really interested in. 

For these reasons, it would be useful to obtain an accurate reduced model
which, though not providing the same detailed information as the kinetic
Wigner-Poisson, could still be able to reproduce the main characteristics  
of quantum plasmas. 

To obtain a set of macroscopic equations for quantum plasmas, first we derive 
a system of reduced `fluid' equations by taking moments of the Wigner-Poisson system. 
Using a Madelung (or eikonal) decomposition, it can be shown that the pressure 
term appearing in the fluid equations can be separated into a classical 
and a quantum part. A working hypothesis is then applied to the
classical term, so as to close the fluid system. Shortly the meaning of classical and quantum contributions to the pressure will be explained. 

This approach to a quantum hydrodynamic model for plasmas appeared in Ref. \cite{Man}. The same quantum fluid model has been applied to several distinct problems in\-vol\-ving charged particle systems, for instance, the nonlinear electron dynamics in thin metal films \cite{Crouseilles}, the excitation of electrostatic wake fields in nanowires \cite{al}, parametric amplification cha\-rac\-te\-ris\-tics in piezoelectric semiconductors \cite{ghosh}, breather waves in semiconductor quantum wells \cite{bre}, multidimensional dissipation-based Schr\"odinger models from quantum Fokker-Planck dynamics \cite{Lop}, the description of quantum diodes in degenerate plasmas \cite{dio} and quantum ion-acoustic waves in single-walled carbon nanotubes \cite{wei}, to name but a few. The extension of the model to incorporate magnetic fields was done in \cite{hs}. 

We take moments of Eq. (\ref{w1}) by integrating
over velocity space. In other words, introducing the standard definitions of density,
mean velocity and pressure 
\begin{equation}
\label{fluid}
n(x,t) = \int f\,dv \,, \quad u(x,t) = \frac{1}{n}\int fv\,dv \,, \quad P(x,t) = 
m\left(\int fv^{2}dv - nu^{2}\right) \, ,
\end{equation}
we get
\begin{eqnarray}
\label{cont}
\frac{\partial\,n}{\partial\,t} + 
\frac{\partial\,(nu)}{\partial\,x} &=& 0 \,, \\ 
\label{force}
\frac{\partial\,u}{\partial\,t} + u\frac{\partial\,u}{\partial\,x} &=& 
\frac{e}{m}\frac{\partial\,\phi}{\partial\,x} - \frac{1}{mn}
\frac{\partial\,P}{\partial\,x} \,.
\end{eqnarray}
A more detailed theory would include the energy transport equation obtained after taking the second-order moment of the Wigner function and the associated time-derivative. 

Equations (\ref{cont}--\ref{force}) 
do not differ from the ordinary evolution equations for a classical fluid.
This may seem strange, but in the following it will appear that the quantum
nature of the system is in fact hidden in the pressure term. Contributions where $\hbar$ explicitly appear can be found only in the higher-order moments. Actually, taking into account the energy transport equation is not sufficient, since in the electrostatic case $\hbar$ appear explicitly only in the equation of motion for the third-order moment \cite{hmbz}. 

To proceed, first notice the equivalence between the Wigner-Poisson and a system of countably many Schr\"odinger equations coupled to the Poisson equation, as has been ma\-the\-ma\-ti\-cal\-ly demonstrated \cite{Markowichb}. More exactly, the reduced one-body Wigner function can always be written as
\begin{equation}
\label{obwf}
f(x,v,t) = \frac{Nm}{2\,\pi\,\hbar}\,\sum_{\alpha=1}^{M}\,p_{\alpha}\int\,ds\,\exp\left(\frac{i\,m\,v\,s}{\hbar}\right)\,\psi_{\alpha}^{*}(x+\frac{s}{2},t)\psi_{\alpha}(x-\frac{s}{2},t) \,, 
\end{equation}
with ensemble probabilities $p_\alpha \geq 0$ so that $\sum_{\alpha=1}^{M}p_\alpha = 1$, for each one-particle ensemble wavefunctions $\psi_{\alpha}(x,t)$ normalized to unity and satisfying  
\begin{equation}
\label{sce}
i\hbar\frac{\partial\psi_\alpha}{\partial t} = - \frac{\hbar^2}{2m}\frac{\partial^2\psi_\alpha}{\partial x^2} - e\phi\psi_\alpha \,, \quad \alpha = 1,...,M \,,
\end{equation}
which is the Schr\"odinger equation for a particle under the action of the mean field electrostatic potential $\phi(x,t)$. In addition, the Poisson equation (\ref{w2}) is rewritten as 
\begin{equation}
\label{poe}
\frac{\partial^2\phi}{\partial x^2} = \frac{e}{\varepsilon_0}\left(N\sum_{\alpha=1}^{M} p_\alpha |\psi_{\alpha}(x,t)|^2 - n_0\right) \,.
\end{equation}
Equations (\ref{sce}--\ref{poe}) constitute the so-called Schr\"odinger-Poisson system, which has to be supplemented with suitable initial and boundary conditions. It provides a way of replacing the original $N-$body problem by a collection of one-body Schr\"odinger equations, coupled by Poisson's equation. From a methodological point of view the Schr\"odinger-Poisson modeling corresponds to put the emphasis again on the wavefunction and not on the (phase space) Wigner function. Collective effects are mediated by the self-consistent potential $\phi$. 

A rigorous proof \cite{Markowichb} of the equivalence of Eqs. (\ref{sce}--\ref{poe}) and the Wigner-Poisson system is beyond the present text. However, at least we can obtain some insight on the interpretation of the ensemble wavefunctions. From Eqs. (\ref{nbody}--\ref{oneb}),
\begin{eqnarray}
f(x_{1},v_{1},t) &=& \int\,dx_{2}dv_{2}...dx_{N}dv_{N}\,f^{N}(x_{1},v_{1},...,x_{N},v_{N},t) \nonumber \\
&=&
N\,\left(\frac{m}{2\,\pi\,\hbar}\right)\sum_{\alpha = 1}^{M}\,p_{\alpha}\,\int\,ds_{1}\,dx_{2}...dx_{N}\exp\left(\frac{i\,m\,v_{1}s_{1}}{\hbar}\right) \times \nonumber \\ 
\label{interm}
&\times& \,\psi^{N\,*}_{\alpha}\left(x_{1}+\frac{s_1}{2},x_{2},...,x_{N},t\right)  
\psi_{\alpha}^{N}\left(x_{1}-\frac{s_1}{2},x_{2},...,x_{N},t\right) \,.
\end{eqnarray}
To be inline with the mean field approximation it is natural to factorize as 
\begin{equation}
\label{ffo}
\psi_{\alpha}^{N}\left(x_{1},x_{2},...,x_{N},t\right) = \psi_{\alpha}(x_1,t) \times ... \times \psi_{\alpha}(x_N,t) \,,
\end{equation}
for the $N-$body wavefunction, fully neglecting correlations. Quantum statistics effects are not taken into account in the \textit{Ansatz} (\ref{ffo}), which does not respect the Pauli principle. With this proviso, inserting Eq. (\ref{ffo}) into Eq. (\ref{interm}) the result is precisely Eq. (\ref{obwf}), with the same statistical weights $p_\alpha$. Hence we can view the one-body ensemble wavefunctions $\psi_{\alpha}(x,t)$ as the result of splitting the $N-$body ensemble wavefunction into the product of identical factors. Actually it not so surprising that a correlationless model could at the end be written in terms of a collection of one-body Schr\"odinger equations. 

Thanks to the Schr\"odinger-Poisson form, we are able to decompose the pressure term in a classical and a quantum part, as follows.
Consider the Wigner distribution in Eq. (\ref{obwf}).  
In terms of the ensemble wavefunctions, from Eq. (\ref{fluid}) one obtains 
\begin{eqnarray}
n &=& N\sum_{\alpha=1}^{M}p_\alpha |\psi_\alpha|^2 \,,\\
nu &=& \frac{i\hbar N}{2m} \sum_{\alpha=1}^{M}p_\alpha (\psi_\alpha \frac{\partial\psi_{\alpha}^{*}}{\partial x} - \psi_{\alpha}^{*} \frac{\partial\psi_{\alpha}}{\partial x}) \,,
\end{eqnarray}
and, after some work,
\begin{eqnarray}
P &=& \frac{N\hbar^2}{4m}  \sum_{\alpha=1}^{M} p_\alpha\left( 2 
\left|\frac{\partial \psi_\alpha}{\partial x}\right|^2 - 
\psi^{*}_\alpha \frac{\partial^2 \psi_\alpha}{\partial x^2} -
\psi_\alpha \frac{\partial^2 \psi^{*}_\alpha}{\partial x^2} \right) 
\nonumber \\ \label{dpr}
&+& ~\frac{N^2\hbar^2}{4mn} \left[ \sum_{\alpha=1}^{M} p_\alpha\left(
\psi^{*}_\alpha \frac{\partial \psi_\alpha}{\partial x} -
\psi_\alpha \frac{\partial \psi^{*}_\alpha}{\partial x} \right)
\right]^2 ~.
\end{eqnarray}

Now we Madelung \cite{mad} decompose the wavefunctions with 
\begin{equation}
\label{psi}
\psi_{\alpha}(x,t) = A_{\alpha}(x,t)\exp{(iS_{\alpha}(x,t)/\hbar)}~,
\end{equation}
where the amplitudes $A_\alpha$ and phases $S_\alpha$ are real functions. We get
\begin{eqnarray}
\label{dns}
n &=& N \sum_{\alpha=1}^{M} p_\alpha A_{\alpha}^2 \,,\\
nu &=& \frac{N}{m} \sum_{\alpha=1}^{M} p_\alpha A_{\alpha}^2 \frac{\partial S_\alpha}{\partial x}
\end{eqnarray}
and also 
\begin{eqnarray}
P = &=& \frac{N^2}{2mn}\sum_{\alpha,\beta=1}^M p_{\alpha}p_{\beta}A_{\alpha}^{2}A_{\beta}^{2}
\left(\frac{\partial S_\alpha}{\partial x}- \frac{\partial S_\beta}{\partial x}\right)^2 \nonumber \\ \label{pppp}
&+& \frac{N\hbar^2}{2m}\sum_{\alpha=1}^{M} p_{\alpha}\left[\left(\frac{
\partial\,A_\alpha}{\partial\,x}\right)^2 - A_{\alpha}\frac{\partial^{2}A_{\alpha}}
{\partial\,x^2}\right] ~.
\end{eqnarray}
In the pressure there is now the explicit presence of $\hbar$. However, notice that since the ensemble wavefunctions satisfy the one-body Schr\"odinger equation (\ref{sce}) both the amplitudes and phases implicitly depend on Planck's constant. 

It is useful to define the kinetic $u_\alpha$ and osmotic $u_{\alpha}^o$ velocities associated to the wavefunction $\psi_\alpha$,
\begin{equation}
u_\alpha = \frac{1}{m}\frac{\partial S_\alpha}{\partial x} \,, \quad u_{\alpha}^o = \frac{\hbar}{m}\frac{\partial A_\alpha/\partial x}{A_\alpha} \,.
\end{equation}
Then it can be verified that the pressure in Eq. (\ref{pppp}) is given by
\begin{equation}
\label{koq}
P = P^k + P^o + P^Q \,,
\end{equation}
where the kinetic pressure $P^k$ is
\begin{equation}
\label{pk}
P^k = \frac{mn}{2}\sum_{\alpha,\beta=1}^{M} \tilde{p}_\alpha \tilde{p}_\beta (u_\alpha - u_\beta)^2 \,,
\end{equation}
the osmotic pressure $P^o$ is 
\begin{equation}
\label{pos}
P^o = \frac{mn}{2}\sum_{\alpha,\beta=1}^{M} \tilde{p}_\alpha \tilde{p}_\beta (u_{\alpha}^o - u_{\beta}^o)^2 \,,
\end{equation}
and the quantum pressure $P^Q$ is
\begin{equation}
\label{pq}
P^Q = - \frac{\hbar^2 n}{4m}\frac{\partial^2}{\partial x^2} \ln n \,.
\end{equation}
In Eqs. (\ref{pk}--\ref{pos}), there is a modified set of ensemble probabilities $\tilde{p}_\alpha = \tilde{p}_{\alpha}(x,t)$,
\begin{equation}
\label{tip}
\tilde{p}_\alpha = \frac{N p_\alpha A_{\alpha}^2}{n} \,.
\end{equation}
The new statistical weights satisfy $\tilde{p}_\alpha \geq 0, \sum_{\alpha=1}^{M} \tilde{p}_\alpha = 1$ as they should. 

For a particular $\psi_\alpha$ the osmotic velocity points to the regions of higher density \cite{bohh, gass}, as is more evident in the three-dimensional version,
\begin{equation}
{\bf u}^{o}_\alpha = (\hbar/m)\nabla\ln A_\alpha \,.
\end{equation}

Both pressures $P^k$ and $P^o$ can be viewed as a measure of the dispersion of velocities, kinetic and osmotic. Consider the following redefined average $<f_\alpha>$ of an ensemble function $f_\alpha = f_{\alpha}(x,t)$:  
\begin{equation}
<f_\alpha> = \sum_{\alpha=1}^M \tilde{p}_\alpha f_\alpha \,.
\end{equation}
Equations (\ref{pk}--\ref{pos}) are equivalent to the standard deviations 
\begin{eqnarray}
\label{flu}
P^k &=&  mn (\langle u_\alpha^2 \rangle - \langle u_\alpha \rangle^2) \,, \\
\label{fluu}
P^o &=&  mn (\langle [u_{\alpha}^{o}]^2 \rangle - \langle u_{\alpha}^{o} \rangle^2) \,.
\end{eqnarray}
For a pure state so that $\tilde{p}_\alpha = \delta_{\alpha\beta}$ for some $\beta$, both $P^k$ and $P^o$ vanishes and only $P^Q$ survives.

Since the kinetic and osmotic pressures are a measure of the kinetic and osmotic velocities dispersion, it is reasonable, although not rigorous, to assume an equation of state so that 
\begin{equation}
\label{hku}
P^k + P^o = P^{C}(n) \,,
\end{equation}
depending only on density. In this way, we obtain 
\begin{equation}
\label{xon}
P = P^{C}(n) - \frac{\hbar^2 n}{4m}\frac{\partial^2}{\partial x^2} \ln n \,.
\end{equation}
For definiteness, we call $P^C$ the ``classical" part of the pressure, in the sense that it represents a measure of the velocities dispersion. However, it explicitly contains Planck's constant since it depends on the osmotic velocities, which always have a purely quantum nature. 

The replacement of the sum of the kinetic and osmotic pressures by a function of the density only requires some comments. Equation (\ref{koq}) is exact but offer no advancement over the Wigner-Poisson formulation, because ultimately it requires the knowledge of the ensemble wavefunctions and then the solution of a countable set of self-consistent Schr\"odinger equations. In classical kinetic theory (therefore, without the osmotic and quantum pressures), it is customary to assume a closure assuming that the standard deviation of the velocities is a function of density only. We have just gone one step further, including also the standard deviation of the osmotic velocities. 

In addition, in the classical limit we expect equations reproducing the classical fluid equations. This is certainly true if $P^k + P^o = P^{C}(n)$ for some appropriate function of density only. Finally, the standard Euler equations are reproduced thanks to the residual classical limit in $P^k$. Or, for $\hbar \equiv 0$, we also have $P^o = 0$ and we expect $P^k$ to be some function of the density only. 

Unlike $P$ in Eq. (\ref{fluid}) which uses the Wigner function, in Eqs. (\ref{flu}--\ref{fluu}) the statistical weights are provided by the $\tilde{p}_\alpha$ in Eq. (\ref{tip}). For a pure state one has $P^C = 0$, which is inline with the understanding that a pure state corresponds to a cold plasma with no dispersion of velocities. The contribution $P^Q$, on the other hand, have no classical counterpart at all. 

It is well-known that the closure problem is a delicate one. The derivation of 
macroscopic models from microscopic models always deserve some degree of approximation
and a more or less phenomenological or ingenuous point of view. The present approach is capable of taking into account the quantum statistics of the 
charge carriers, represented by an appropriated equation of state. For example, assuming interaction with a heat bath an isothermal equation of state is indicated. For fast phenomena where thermal relaxation has no time to occur, an adiabatic equation of state can be used. In addition, Maxwell-Boltzmann or Fermi-Dirac distributions can be applied according resp. to the dilute or dense class of plasma. Moreover, the model takes into account quantum diffraction effects, in particular tunneling and wave packet dispersion, present in the quantum part $P^Q$ of the pressure. It is also able to reproduce the linear dispersion relation from kinetic theory if the equation of state is adequate, except for purely kinetic phenomena. Finally, the quantum fluid model reduces to the standard Euler equations in the formal classical limit and are sufficiently simple to be amenable to efficient numerical simulation.   

The proposed simplification becomes more justified for an important class of statistical ensembles, where the wavefunctions have all equal (but not necessarily constant) amplitude, 
\begin{equation}
\label{ghk}
\psi_\alpha = \sqrt{\frac{n}{N}}\,\,e^{iS_{\alpha}/\hbar} \,.
\end{equation}
Then the osmotic pressure identically vanishes and one has $\tilde{p}_\alpha = p_\alpha$, so that Eq. (\ref{flu}) becomes the usual standard deviation of the kinetic velocities. Hence $P^k$ can be interpreted in full analogy with the standard thermodynamic pressure. The velocities dispersion arises just from the randomness of the phases of the wavefunctions. The approximation can be viewed as a first step beyond
the standard homogeneous equilibrium of a fermion gas, for which each state
can be represented by a plane wave
\begin{equation}
\psi_{\alpha}(x,t) = A_0 \exp{(i m u_\alpha x/\hbar})\,,
\end{equation}
with the amplitude $A_0$ and 
the velocities $u_\alpha$ spatially constant.
In the generalization (\ref{ghk}), both the amplitude and the velocity
can be spatially modulated, although the amplitude is the same for all states.
For systems not so far from equilibrium, this appears to be reasonable. 

With the hypothesis (\ref{hku}) the force equation 
$(\ref{force})$ can be written as
\begin{equation}
\label{force1}
\frac{\partial\,u}{\partial\,t} + u\frac{\partial\,u}{\partial\,x} = - 
\frac{1}{mn}\frac{\partial\,P^{C}(n)}{\partial\,x} + 
\frac{e}{m}\frac{\partial\,\phi}{\partial\,x}  -
\frac{1}{mn}\frac{\partial\,P^Q}{\partial\,x} \,.
\end{equation}
Using the identity
\begin{equation}
\frac{1}{mn}\frac{\partial\,P^Q}{\partial\,x} =
- \frac{\hbar^2}{2m^2}
\frac{\partial}{\partial x}\left(\frac{\partial^{2}(\sqrt{n})/\partial\,x^2}
{\sqrt{n}}\right)~,
\end{equation}
we can rewrite the basic quantum hydrodynamic model for plasmas in terms of the continuity equation (\ref{cont}) 
and the force equation 
\begin{equation}
\label{force2}
\frac{\partial\,u}{\partial\,t} + u\frac{\partial\,u}{\partial\,x} = - 
\frac{1}{mn}\frac{\partial\,P^{C}(n)}{\partial\,x} + 
\frac{e}{m}\frac{\partial\,\phi}{\partial\,x} + \frac{\hbar^2}{2m^2}
\frac{\partial}{\partial x}\left(\frac{\partial^{2}(\sqrt{n})/\partial\,x^2}
{\sqrt{n}}\right) \,.
\end{equation}
In the limit $\hbar \rightarrow 0$ this is formally equal to Euler's equation for an electron fluid in the presence of an electric field $- \partial\phi/\partial x$. Finally, we have the Poisson equation 
\begin{equation}
\label{pois2}
\frac{\partial^{2}\phi}{\partial\,x^2} = \frac{e}{\varepsilon_0}(n - 
n_{0}) ~,
\end{equation}
The only difference to classical plasmas is the $\sim \hbar^2$ term in Eq. (\ref{force2}), the so-called Bohm potential term. While mathematically the Bohm potential is equivalent to a pressure to be inserted in the momentum transport equation, physically it corresponds to typical quantum phenomena like tunneling and wave packet spreading. Therefore it is not a pressure in the thermodynamic sense. Besides, it survives even for a one-particle pure state system.

Many quantum hydrodynamical models are popular in the context of semiconductor physics. For instance Gardner \cite{Gard} considered a quantum corrected displaced Maxwellian as introduced by Wigner \cite{Wign}. More exactly, it is possible to find the leading quantum correction $f_{1}(x,v,t)$ for a momentum-shifted local Maxwell-Boltzmann equilibrium $f_{0}(x,v,t)$, setting $f(x,v,t) = f_{0}(x,v,t) + \hbar^2 f_{1}(x,v,t)$ in the semiclassical quantum Vlasov equation (\ref{w4}) and collecting equal powers of $\hbar^2$. In other words set 
\begin{equation}
f_{0}(x,v,t) = n(x,t) \left(\frac{m}{2\pi\kappa_B T(x,t)}\right)^{1/2} \exp\left(\frac{- m [v - u(x,t)]^2}{2\kappa_B T(x,t)}\right) \,,
\end{equation}
on the assumption of a non-degenerate quasi-equilibrium state. After calculating $f_1$, insert $f = f_0 + \hbar^2 f_1$ in the pressure in Eq. (\ref{fluid}). A quantum hydrodynamical model similar to Eqs. (\ref{cont}) and (\ref{force2}--\ref{pois2}) is then found (and generalized by the inclusion of an energy transport equation). By definition, the resulting system is res\-tric\-ted to quasi-Maxwellian, dilute systems. We also observe that in the case of a self-consistent problem the electrostatic potential should have been also expanded in powers of some non-dimensional quantum parameter, which was not done in \cite{Gard}. The conclusion is that the procedures involving a Madelung decomposition of the quantum ensemble wavefunctions or a quantum corrected Wigner function equilibrium involve working hypotheses which are not rigorously justified. Starting from the Wigner-Poisson system, which is by definition collisionless, hardly one could derive rigorous macroscopic theories relying on quasi-equilibrium assumptions. Nevertheless, the numerical and analytical advantages of quantum fluid models over quantum kinetic models make them much more popular than the kinetic treatment, to approach the nonlinear aspects of quantum plasma physics. 

Alternatively, general thermodynamic arguments by Ancona and Tiersten have been used to derive the Bohm potential. Ref. \cite{Tiersten} argues that the internal e\-ner\-gy of the electron fluid in an electron-hole semiconductor should depend not only on the density but on the density gradient too, in order to extend the standard drift-diffusion model so as to include the quantum-mechanical behavior exhibited in strong inversion layers. Their method defines a ``double-force" and a ``double-pressure vector" which allows for changes of the internal e\-ner\-gy of the electron gas purely due to density fluctuations. Postulating a linear dependence of the double-pressure vector on density gradients (see Eq. (3.3) of \cite{Tiersten}) and working out the conservation laws of charge, mass, linear momentum and energy, Ancona and Tiersten found a ge\-ne\-ra\-li\-zed chemical potential composed of two contributions: (a) a gradient-independent term which can be modeled by the equation of state of a zero-temperature Fermi gas or any other appropriated form. This corresponds to the classical pressure $P^{C}(n)$ of the quantum hydrodynamical model for plasmas; (b) a Bohm potential term proportional to a phenomenological parameter left initially free. 

Later, Ancona and Iafrate \cite{Anco} obtained the expression of the phenomenological coefficient of Ref. \cite{Tiersten}. In \cite{Anco}, which is similar to \cite{Gard}, the first order quantum correction for a Maxwell-Boltzmann equilibrium was found from the semiclassical quantum Vlasov equation and then employed to calculate the particle density and the stress tensor. Eliminating the potential function between the two expressions, an equation of state relating the stress tensor and the particle density and gradient was derived, containing the Bohm potential. The gradient dependence of the stress tensor arises because of the quantum mechanical nonlocality. Once again, the demonstration in \cite{Anco} is valid when quantum contributions to the self-consistent mean field potential can be ignored: no expansion of $\phi$ in powers of a quantum parameter was used. Moreover, the dilute and semiclassical si\-tua\-tion is supposed, where the Maxwell-Boltzmann statistics apply and a small dimensionless quantum parameter exist. 

Further, quantum hydrodynamic models with a ``smoothed" potential have been derived, to handle discontinuities in potential barriers in semiconductors \cite{gari}. Energy transport \cite{dego} and quantum drift-diffusion \cite{anc} have been also employed to model transport in ultra-small electronic devices, including the addition of viscosity effects \cite{jun}. 

It is amazing to notice how quantum hydrodynamic models are widespread in several areas, without apparently the explicit knowledge of the different communities. For instance, in molecular physics, the simplest gradient functional theory is given by the
Thomas-Fermi-Dirac-von Weizs\"acker functional, which provides a suitable expression for the energy density of molecules \cite{Chan}. It is composed by a Thomas-Fermi term for the uniform electron gas \cite{Thomas, Fermi}, a Dirac exchange functional \cite{Dirac}, the Coulomb term for the electron gas, a nuclear attraction term and the von Weizs\"acker term \cite{weiss}, with an adjustable fitting parameter. With this phenomenological parameter set to unity, the von Weiss\"acker term gives exactly the Bohm potential. 

It is useful to remember the obvious limitations of the quantum hydrodynamic model for plasmas: (a) since it is a fluid model, it is applicable only for long wavelengths. This requirement is expressed roughly as $\lambda > \lambda_D$ for non-degenerate and $\lambda > \lambda_F$ for degenerate systems, where $\lambda$ is the pertinent wavelength. Kinetic phenomena (Landau damping, the plasma echo \cite{ech}) requiring a detailed knowledge of the equilibrium Wigner function certainly deserve a kinetic treatment; (b) no energy transport equation was included. This limitation is not definitive since it is a simple exercise to calculate it taking the second-order moment of the quantum Vlasov equation; (c) it applies only to non-relativistic phenomena; (d) no spin effects are included, except for the equation of state which can, in a certain measure, represent some quantum statistical effects. It suffices to apply the equation of state for a dense, degenerate electron gas. In the same way, some relativistic effects can be also incorporated with an equation of state for a relativistic electron gas; (e) no magnetic fields are in the present formulation; (f) the sum of kinetic and osmotic pressures is assumed to be a function of the density only. This is in the spirit of density functional theories and appears reasonable in the vicinity of homogeneous equilibria, but may prove inappropriate in far from equilibrium situations; (g) strongly coupled plasmas are out of reach since the starting point for the derivation was the quantum Vlasov equation, which is valid if correlations are not so important. 

It is remarkable that the inclusion of magnetic fields \cite{hs} starting from the Wigner-Maxwell system can also be made in terms of the eikonal decomposition method, with the resulting quantum hydrodynamic model having just the addition of a magnetic $\sim {\bf u}\times{\bf B}$ force. 

\section{Concluding remarks}
Quantum plasma theory started in the 50's with the analysis of the ground state and correlation energies of the dense electron gas in metals \cite{dubo, gell, Sawada}, through quantum field-theoretical techniques. Pioneering works \cite{Lind, Klim, Pin} also focused on collective oscillation modes in dense plasmas, described by the collisionless quantum Boltzmann equation. After a more or less quiescent period, in the last decade we observe a new wave of interest on quantum plasmas. Besides applications to diverse systems ranging from nanoscale electronic devices and dense astrophysics environments to intense laser-solid density plasma interaction experiments, an important step justifying the renewed attention to quantum plasmas is the emergence of efficient macroscopic models. This short introduction has been essentially dedicated to the analysis of the transition from the microscopic (kinetic) to macroscopic (fluid) modeling of quantum plasmas.

Quantum plasma physics is a rapidly growing field, with unpredictable consequences. Below we list other fundamental topics for brevity not discussed here, currently being scrutinized in the area.
\begin{itemize}
\item The dynamics of spin degrees of freedom and the underlying ferromagnetic behavior in quantum plasmas \cite{m2}--\cite{zama}.
\item Relativistic quantum plasma models \cite{mel}--\cite{tito}, an essential subject since frequently quantum and relativistic effects appear simultaneously as e.g. in applications to compact astrophysical objects and intense laser plasmas. 
\item Extensions to include exchange-correlation and collisional contributions. If the anti-symmetric nature of the $N-$body wavefunction is considered a Hartree-Fock term appear in the interaction energy \cite{Klim}. However, frequently such nonlocal exchange-correlation terms are replaced by local phenomenological expressions, along with the adiabatic local density approximation \cite{Guse, Sant}. Also, we mention the Wigner-Fokker-Planck system approach to model quantum dissipation \cite{Lop, Arn}. 
\end{itemize}

Least but not last, in this introductory text we have focused on the detailed {\it derivation} of the basic kinetic and hydrodynamic models for quantum plasmas, in such a way to be accessible to beginners. A more complete account on the present {\it applications} of such quantum plasma models can be found in \cite{Haas}--\cite{Manfredi}. 

\begin{acknowledgments}
This work was supported by CNPq (Conselho Nacional de Desenvolvimento Cient\'{\i}fico e Tecnol\'ogico).
\end{acknowledgments}


\begin{thebibliography}{99} 
\bibitem{Haas} F. Haas. \textit{Quantum Plasmas: an Hydrodynamic Approach}, (Springer, New York, 2011, in press).
\bibitem{Shukla1} P. K. Shukla and B. Eliasson. Rev. Mod. Phys. (2011, in press).
\bibitem{Shukla2} P. K. Shukla and B. Eliasson. Phys. Uspekhi \textbf{53}, 55 (2010). 
\bibitem{Manfredi} G. Manfredi. Fields Inst. Commun. \textbf{46}, 263 (2005).
\bibitem{Chabrier} G. Chabrier, F. Douchin and A. Y. Potekhin. J. Phys.: Condens. Matter \textbf{14}, 9133 (2002).
\bibitem{Marklund} M. Marklund and P. K. Shukla. Rev. Mod. Phys. \textbf{78}, 591 (2006).
\bibitem{Marko} P. A. Markowich, C. A., Ringhofer and C. Schmeiser. \textit{Semiconductor Equations}, (Springer, Wien, 1990).
\bibitem{Jungel} A. J\"ungel. \textit{Transport Equations for Semiconductors}, (Springer, Berlin-Heidelberg, 2009).
\bibitem{Crouseilles} N. Crouseilles, P. A. Hervieux and G. Manfredi. Phys. Rev. B \textbf{78}, 155412 (2008). 
\bibitem{Glenzer} S. H. Glenzer and R. Redmer. Rev. Mod. Phys. \textbf{81}, 1625 (2009).
\bibitem{Haas2} F. Haas, G. Manfredi and M. Feix. Phys. Rev. E \textbf{62}, 2763 (2000). 
\bibitem{Gregori} G. Gregori and D. O. Gericke. Phys. Plasmas \textbf{16}, 056306 (2009).
\bibitem{Salinas} S. R. A. Salinas. \textit{Introduction to Statistical Physics}, (Springer-Verlag, New York, 2001).
\bibitem{fren} W. R. Frensley. Rev. Mod. Phys. \textbf{62}, 745 (1990).
\bibitem{Murillo} M. S. Murillo. J. Phys. A: Math. Theor. \textbf{42}, 214054 (2009). 
\bibitem{Friedel} J. Friedel. Adv. Phys. \textbf{3}, 446 (1954).
\bibitem{Lind} J. Lindhard. Dan. Vidensk. Selsk., Mat. Fys. Medd. \textbf{28}, 1 (1954).
\bibitem{Bonitz} M. Bonitz, D. Semkat, A. Filinov, V. Golubnychyi, D. Kremp, D. O. Gericke, M. S. Murillo, V. Filinov, V. Fortov, W. Hoyer and S. W. Koch. J. Phys. A: Math. Gen. \textbf{36}, 5921 (2003). 
\bibitem{Wign} E. Wigner. Phys. Rev. \textbf{40}, 749 (1932). 
\bibitem{Klim} Y. Klimontovich and V. P. Silin. In: Drummond, J. E. (ed.) \textit{Plasma Physics}, pp. 35--87, (McGraw-Hill, New York, 1961). 
\bibitem{Bogoliubovb} N. N. Bogoliubov. J. Exp. Theor. Phys. \textbf{16}, 691 (1946).
\bibitem{Born} M. Born and H. S. Green. Proc. Roy. Soc. A \textbf{188}, 10 (1946).
\bibitem{Kirkwoodb} J. G.  Kirkwood. J. Chem. Phys. \textbf{14}, 180 (1946).
\bibitem{Yvon} J. Yvon. \textit{La Th\'eorie Statistique des Fluides}, (Hermann, Paris, 1935).
\bibitem{DiVentra} M. Di Ventra. \textit{Electrical Transport in Nanoscale Systems}, (Cambridge, New York, 2008).
\bibitem{Haug} H. J. W. Haug and A. P. Jauho. \textit{Quantum Kinetics in Transport and Optics of Semiconductors}, (Springer-Verlag, Berlin-Heidelberg, 2008).
\bibitem{Kadanoff} L. P. Kadanoff and G. Baym. \textit{Quantum Statistical Mechanics: Green's Function Methods in Equilibrium and Non-Equilibrium Problems}, (Benjamin, New York, 1962).
\bibitem{Hillery} M. Hillery, R. F. O'Connell, M. O. Scully and E. P. Wigner. Phys. Rep. \textbf{106}, 121 (1990).
\bibitem{Kluk} N. C. Kluksdahl, A. M. Kriman, D. K. Ferry and C. Ringhofer. Phys. Rev. B. \textbf{39}, 7720 (1989).
\bibitem{Man} G. Manfredi and F. Haas. Phys. Rev. B \textbf{64}, 075316 (2001).
\bibitem{al} S. Ali, H. Ter\c{c}as and J. T. Mendon\c{c}a. Phys. Rev. B \textbf{83}, 153401 (2011).
\bibitem{ghosh} S. Ghosh, S. Dubey and R. Vanshpal. Phys. Lett. A \textbf{375}, 43 (2010).
\bibitem{bre} F. Haas, G. Manfredi, P. K. Shukla and P.-A. Hervieux. Phys. Rev. B \textbf{80}, 073301 (2009).
\bibitem{Lop} J. L. L\'opez. Phys. Rev. E \textbf{69}, 026110 (2004).
\bibitem{dio} P. K. Shukla and B. Eliasson. Phys. Rev. Lett. \textbf{100}, 036801 (2008).
\bibitem{wei} L. Wei and Y. N. Wang. Phys. Rev. B \textbf{75}, 193407 (2007). 
\bibitem{hs} F. Haas. Phys. Plasmas \textbf{12}, 062117 (2005).
\bibitem{hmbz} F. Haas, M. Marklund, G. Brodin and J. Zamanian. Phys. Lett. A \textbf{374}, 481 (2010). 
\bibitem{Markowichb} P. A. Markowich. Math. Meth. Appl. Sci. \textbf{11}, 459 (1989). 
\bibitem{mad} E. Madelung. Phys. \textbf{40}, 332 (1926).
\bibitem{bohh} D. Bohm and B. J. Hiley. \textit{The Undivided Universe: an Ontological Interpretation of Quantum Theory}, (Routledge, London, 1993).
\bibitem{gass} I. Gasser, C. Lin and P. A. Markowich. Taiwan. J. Math. \textbf{4}, 501 (2000).
\bibitem{Gard} C. L. Gardner. SIAM J. Appl. Math. \textbf{54}, 409 (1994).
\bibitem{Tiersten} M. G. Ancona and H. F. Tiersten. Phys. Rev. B. \textbf{35}, 7959 (1987).
\bibitem{Anco} M. G. and G. J. Iafrate. Phys. Rev. B \textbf{39}, 9536 (1989).
\bibitem{gari} C. L. Gardner and C. Ringhofer. Phys. Rev. E \textbf{393}, 157 (1996). 
\bibitem{dego} P. Degond, F. M\'ehats and C. Ringhofer. J. Stat. Phys. \textbf{118}, 625 (2005).
\bibitem{anc} M. Ancona. COMPEL \textbf{6}, 11 (1987).
\bibitem{jun} A. J\"ungel and S. Tang. Appl. Num. Math. \textbf{56}, 899 (2006).
\bibitem{Chan} G. K. Chan, A. J. Cohen and N. C. Handy. J. Chem. Phys. \textbf{114}, 631 (2001). 
\bibitem{Thomas} L. H. Thomas. Proc. Cambridge Philos. Soc. \textbf{26}, 376 (1930).
\bibitem{Fermi} E. Fermi. Z. Phys. \textbf{48}, 73 (1928). 
\bibitem{Dirac} P. A. M. Dirac. Proc. Cambridge Philos. Soc. \textbf{26}, 376 (1930).
\bibitem{weiss} C. F. Weizs\"acker. Z. Phys. \textbf{96}, 431 (1935).
\bibitem{ech} G. Manfredi and M. R. Feix. Phys. Rev. E \textbf{53}, 6460 (1996). 
\bibitem{dubo} D. F. Dubois. Ann. Phys. \textbf{7}, 174 (1959).
\bibitem{gell} M. Gellmann and K. A. Brueckner. Phys. Rev. \textbf{106}, 364 (1957). 
\bibitem{Sawada} K. Sawada. Phys. Rev. \textbf{106}, 372 (1957).
\bibitem{Pin} D. Pines and P. Nozi\`eres. \textit{The Theory of Quantum Liquids} (New York,  W. A. Benjamin, 1966). 
\bibitem{m2} M. Marklund and G. Brodin. Phys. Rev. Lett. \textbf{98}, 025001 (2007).
\bibitem{m1} G. Brodin and M. Marklund. New J. Phys. \textbf{9}, 277 (2007).
\bibitem{shuu} P. K. Shukla. Nature Phys. \textbf{5}, 92 (2009).
\bibitem{m4} G. Brodin, A. P. Misra and M. Marklund. Phys. Rev. Lett. \textbf{105}, 105004 (2010).
\bibitem{zama} J. Zamanian, M. Marklund and G. Brodin. New J. Phys. \textbf{12}, 043019 (2010).
\bibitem{mel} D. B. Melrose. {\it Quantum Plasmadynamics: Unmagnetized Plasmas}, Lecture Notes in Physics Vol. 735 (Springer, New York, 2008).
\bibitem{Zhu} J. Zhu and P. Ji. Phys. Rev. E \textbf{81}, 036406 (2010). 
\bibitem{ase} F. A. Asenjo, V. Munoz, J. A. Valdivia and S. M. Mahajan. 
Phys. Plasmas \textbf{18}, 012107 (2011). 
\bibitem{tito} J. T. Mendon\c{c}a. Phys. Plasmas \textbf{18}, 062101 (2011).
\bibitem{Guse} G. M. Gusev, A. A. Quivy, T. E. Laman, J. R. Leite, A. K. Bakarov, A. I. Topov, O. Estibals and J. C. Portal. Phys. Rev. B \textbf{65}, 205316 (2002). 
\bibitem{Sant} M. Santer, B. Mehlig and M. Moseler. Phys. Rev. Lett. \textbf{89}, 266801 (2002).
\bibitem{Arn} A. Arnold, J. L. L\'opez, P. Markowich and J. Soler. Rev. Mat. Iberoamericana \textbf{20}, 771 (2004).
\end{thebibliography}
\end{document}